\documentstyle[12pt]{article}
\begin{document}

\title{\textbf{Thermodynamics of apparent horizon in modified FRW universe with
power-law corrected entropy}}

\author{K. Karami$^{1,2}$\thanks{E-mail: KKarami@uok.ac.ir} , A. Abdolmaleki$^{1}$, N. Sahraei$^{1}$, S. Ghaffari$^{1}$\\\\
$^{1}$\small{Department of Physics, University of Kurdistan,
Pasdaran St., Sanandaj, Iran}\\$^{2}$\small{Research Institute for
Astronomy $\&$ Astrophysics of Maragha (RIAAM), Maragha, Iran}\\
}

\maketitle

\begin{abstract}
We derive the modified Friedmann equation corresponding to the
power-law corrected entropy-area relation $S_{\rm
A}=\frac{A}{4}\left[1-K_{\alpha} A^{1-\frac{\alpha}{2}}\right]$
which is motivated by the entanglement of quantum fields in and out
of the apparent horizon. We consider a non-flat modified FRW
universe containing an interacting viscous dark energy with dark
matter and radiation. For the selected model, we study the effect of
the power-law correction term to the entropy on the dynamics of dark
energy. Furthermore, we investigate the validity of the generalized
second law (GSL) of gravitational thermodynamics on the apparent
horizon and conclude that the GSL is satisfied for $\alpha<2$.
\end{abstract}

%\noindent{\textbf{PACS numbers:}~~~95.36.+x, 04.60.-m }\\

\newpage
%-----------------------------------------------------------------------------------------------
\section{Introduction}
The present acceleration of the universe expansion has been well
established through numerous and complementary cosmological
observations \cite{Riess}. One explanation for the cosmic
acceleration is the dark energy (DE), an exotic energy with negative
pressure. Although the nature and cosmological origin of DE is still
enigmatic at present, a great variety of models have been proposed
to describe the DE (for review see \cite{Padmanabhan}). One of
interesting issues in modern cosmology is the thermodynamical
description of the accelerating universe driven by the DE.

In black hole physics, it was found that black holes emit Hawking
radiation with a temperature proportional to their surface gravity
at the event horizon and they have an entropy which is one quarter
of the area of the event horizon \cite{Hawking}. The temperature,
entropy and mass of black holes satisfy the first law of
thermodynamics \cite{Bardeen}. It was shown that the Einstein
equation can be derived from the first law of thermodynamics by
assuming the proportionality of entropy and the horizon area
\cite{Jacobson}.

The relation between the Einstein equation and the first law of
thermodynamics has been generalized to the cosmological context. It
was shown that by applying the Clausius relation $-{\rm d}E=T_A{\rm
d} S_A$ to the apparent horizon $\tilde{r}_{\rm A}$, the Friedmann
equation in the Einstein gravity can be derived if we take the
Hawking temperature $T_{\rm A}=1/(2\pi \tilde{r}_{\rm A})$ and the
entropy $S_{\rm A}=A/4$ on the apparent horizon, where $A$ is the
area of the horizon \cite{Cai05}. The equivalence between the first
law of thermodynamics and Friedmann equation was also found for
gravity with Gauss-Bonnet term, the Lovelock gravity theory and the
braneworld scenarios \cite{Cai05,Akbar,Sheykhi1}.

Note that in thermodynamics of apparent horizon in the standard
Friedmann-Robertson-Walker (FRW) cosmology, the geometric entropy is
assumed to be proportional to its horizon area, $S_A={A}/{4}$
\cite{Cai05}. However, this definition for the entropy can be
modified from the inclusion of quantum effects. For instance in
quantum tunneling formalism, taking into account the quantum back
reaction effects in the spacetime found by conformal field theory
methods and using the second law of thermodynamics, the corrections
to the both semiclassical Bekenstein-Hawking area law ($S_{\rm
BH}=A/4$) and Friedmann equation can be obtained \cite{Banerjee}.
The quantum corrections provided to the entropy-area relationship
lead to the curvature correction in the Einstein-Hilbert action and
vice versa \cite{Suj}. The power-law quantum correction to the
horizon entropy motivated by the entanglement of quantum fields
between inside and outside of the horizon is given by \cite{Saurya}
\begin{equation}
S_{\rm A}=\frac{A}{4}\left[1-K_{\alpha}
A^{1-\frac{\alpha}{2}}\right],\label{ec}
\end{equation}
where we take $c=k_B=\hbar=G=1$. Also $\alpha$ is a dimensionless
parameter and
\begin{equation}
K_\alpha=\frac{\alpha(4\pi)^{\frac{\alpha}{2}-1}}{(4-\alpha)r_c^{2-\alpha}},
\end{equation}
where $r_c$ is the crossover scale. Note that in the case of
$\alpha=0=K_{\alpha}$, Eq. (\ref{ec}) reduces to the well-known
Bekenstein-Hawking entropy-area relation $S_{\rm A}=S_{\rm BH}=A/4$.

Besides the first law of thermodynamics, a lot of attention has been
paid to the generalized second law (GSL) of thermodynamics in the
accelerating universe driven by the DE
\cite{Izquierdo1,Sadjadi07,Zhou07,Gong07,Sheykhi2,Karami1,Radicella}.
The GSL of thermodynamics like the first law is an accepted
principle in physics. According to the GSL, the entropy of matter
inside the horizon plus the entropy of the horizon do not decrease
with time \cite{Bekenstein,Davies,Izquierdo2,Wang1}.

Here, we would like to examine whether the power-law corrected
entropy (\ref{ec}) together with the matter field entropy inside the
apparent horizon will satisfy the GSL of thermodynamics. To be more
general we will consider an interacting viscous DE with dark matter
(DM) and radiation. The observations indicate that the universe
media is not a perfect fluid and the viscosity is concerned in the
evolution of the universe (see \cite{Ren1} and references therein).

This paper is organized as follows. In section 2, using the Clausius
relation we derive the modified Friedmann equation corresponding to
the power-law corrected entropy (\ref{ec}). In section 3, we study
the interacting viscous DE with DM and radiation in a non-flat
modified FRW universe. In section 4, we investigate the effect of
the power-law correction term to the entropy on the dynamics of DE.
Section 5 is devoted to conclusions. In Appendix A, we investigate
the validity of the GSL of gravitational thermodynamics with
power-law corrected entropy for the universe enclosed by the
apparent horizon.

%-----------------------------------------------------------------------------------------------
\section{Clausius relation and modified Friedmann equation}
In the framework of FRW metric,
\begin{equation}
{\rm d}s^2 = h_{ij}{\rm d}x^i{\rm d}x^j + \tilde{r}^2{\rm
d}\Omega^2,
\end{equation}
where $\tilde{r}(t) = a(t)r$, $x^i = (t, r)$ and $h_{ij}$ =
diag($-1, a^2/(1 - kr^2)$), $i,j=0,1$, by setting
\begin{equation}
f:=h^{ij}\partial_{i}\tilde{r}\partial_{j}\tilde{r}=1-\left(H^2+\frac{k}{a^2}\right)\tilde{r}^2=0,\label{f}
\end{equation}
the location of the apparent horizon in the FRW universe is
obtained as \cite{Poisson}
\begin{equation}
\tilde{r}_{\rm A}=H^{-1}(1+\Omega_k)^{-1/2}.\label{ra}
\end{equation}
Here $\Omega_{k}=k/(a^2H^2)$ and $k=0,1,-1$ represent a flat, closed
and open universe, respectively. The Hawking temperature on the
apparent horizon is given by \cite{Cai05}
\begin{equation}
T_{\rm A}=\frac{1}{2\pi \tilde{r}_{\rm
A}}\left(1-\frac{\dot{\tilde{r}}_{\rm A}}{2H\tilde{r}_{\rm A}}
\right),\label{TA1}
\end{equation}
where $\frac{\dot{\tilde{r}}_{\rm A}}{2H\tilde{r}_{\rm A}}<1$
ensure that the temperature is positive.

To derive the modified Friedmann equation corresponding to the
power-law corrected entropy (\ref{ec}) we start with the Clausius
relation \cite{Jacobson}
\begin{equation}
-{\rm d}E=T_{\rm A}{\rm d}S_{\rm A},
\end{equation}
where $-{\rm d}E$ is the amount of energy crossing the apparent
horizon during the infinitesimal time interval ${\rm d}t$ in which
the radius of the apparent horizon is assumed to be fixed, i.e.
$\dot{\tilde{r}}_{\rm A} = 0$ \cite{Cai09}. This yields $T_{\rm
A}=1/(2\pi \tilde{r}_{\rm A})$.

Following \cite{Cai5} we have
\begin{equation}
-{\rm d}E=4\pi\tilde{r}_{\rm A}^3(\rho+p)H{\rm d}t,\label{dE}
\end{equation}
where $\rho$ and $p$ are the energy density and pressure of the
fluid, respectively, inside the universe and satisfy the energy
conservation law
\begin{equation}
\dot{\rho}+3H(\rho+p)=0.\label{econs}
\end{equation}
From Eqs. (\ref{ec}), (\ref{ra}), (\ref{TA1}) and using
$A=4\pi\tilde{r}_{\rm A}^2$ one can obtain
\begin{equation}
T_{\rm A}{\rm d}S_{\rm A}=T_{\rm A}\frac{\partial S_{\rm
A}}{\partial A}~{\rm
d}A=-\left(\dot{H}-\frac{k}{a^2}\right)\left\{1-\frac{\alpha}{2}\left[\left(H^2+\frac{k}{a^2}\right)^{1/2}r_c\right]^{\alpha-2}\right\}\tilde{r}_{\rm
A}^3H{\rm d}t.\label{TdS}
\end{equation}
Equating (\ref{dE}) with (\ref{TdS}) and using (\ref{econs}) gives
\begin{equation}
2H\left(\dot{H}-\frac{k}{a^2}\right)\left\{1-\frac{\alpha}{2}
\left[\left(H^2+\frac{k}{a^2}\right)^{1/2}r_c\right]^{\alpha-2}\right\}=\frac{8\pi}{3}\dot{\rho}.\label{feq1}
\end{equation}
Integrating with respect to cosmic time $t$ we get the modified
Friedmann equation
\begin{equation}
H^2+\frac{k}{a^2}-r_c^{-2}\left[r_c^{\alpha}\left(H^2+\frac{k}{a^2}\right)^{\alpha/2}-1\right]=\frac{8\pi}{3}\rho,
\label{FRW1}
\end{equation}
which in the absence of correction term, i.e. $\alpha=0$, it
recovers the well-known first Friedmann equation in the standard FRW
cosmology.
%-----------------------------------------------------------------------------------------------

\section{Interacting viscous DE, DM and radiation}
Here we consider a non-flat FRW universe containing the DE, DM and
radiation. Hence the first modified Friedmann equation (\ref{FRW1})
corresponding to the power-law corrected entropy (\ref{ec}) takes
the form
\begin{equation}
H^2+\frac{k}{a^2}-r_c^{-2}\left[r_c^{\alpha}\left(H^2+\frac{k}{a^2}\right)^{\alpha/2}-1\right]
=\frac{8\pi}{3}(\rho_D+\rho_m+\rho_r),\label{eqf1}
\end{equation}
where $\rho_D$, $\rho_m$ and $\rho_r$ are the energy density of
DE, DM and radiation, respectively.

Using the following definitions
\begin{equation}
\Omega_{D}=\frac{8\pi \rho_{D}}{3H^2},~~~\Omega_{m}=\frac{8\pi
\rho_{m}}{3H^2},~~~\Omega_{r}=\frac{8\pi
\rho_{r}}{3H^2},\label{Omega1}
\end{equation}
\begin{eqnarray}
\Omega_\alpha&=&(Hr_c)^{-2}\left[(Hr_c)^{\alpha}(1+\Omega_{k})^{\alpha/2}-1\right],
\nonumber\\
&=&(1+\Omega_k)\left(\frac{\tilde{r}_A}{r_c}\right)^2\left[\left(\frac{\tilde{r}_A}{r_c}\right)^{-\alpha}-1\right]
,\label{Omega2}
\end{eqnarray}
one can rewrite Eq. (\ref{eqf1}) as
\begin{equation}
1+\Omega_k=\Omega_D+\Omega_m+\Omega_r+\Omega_\alpha.\label{eqf2}
\end{equation}
Here, we consider a viscous model of DE. In an isotropic and
homogeneous FRW universe, the dissipative effects arise due to the
presence of bulk viscosity in cosmic fluids. The DE with bulk
viscosity has a peculiar property to cause accelerated expansion of
phantom type in the late evolution of the universe
\cite{Brevik,Ren2}. Note that the total energy density still
satisfies the conservation law (\ref{econs}) where
\begin{equation}
\rho=\rho_D+\rho_m+\rho_r,\label{rho}
\end{equation}
\begin{equation}
p=\tilde{p}_D+p_r,\label{p1}
\end{equation}
and
\begin{equation}
\tilde{p}_D=p_D-3H\xi,
\end{equation}
is the effective pressure of the DE and $\xi$ is the bulk viscosity
coefficient \cite{Ren1,Ren2}. Note that $p_r=\rho_r/3$ and the DM is
pressureless, i.e. $p_m=0$. Here like \cite{Sheykhi3}, if we assume
$\xi=\varepsilon\rho_D H^{-1}$, where $\varepsilon$ is a constant
parameter, then the total pressure yields
\begin{equation}
p=(\omega_D-3\varepsilon)\rho_D+\frac{1}{3}\rho_r,\label{p2}
\end{equation}
where $\omega_D=p_D/\rho_D$ is the equation of state (EoS) parameter
of the viscous DE.

We further assume that the viscous DE, DM and radiation interact
with each other. Recently the scenario in which the DE interacts
with DM and radiation has been introduced to resolve the cosmic
triple coincidence problem \cite{triple}. In the presence of
interaction, the energy conservation laws for the viscous DE, DM and
radiation are not separately hold and we have
\begin{equation}
\dot{\rho_D}+3H\rho_D(1+\omega_D)=9H^2\xi-Q,\label{continD}
\end{equation}
\begin{equation}
\dot{\rho_m}+3H\rho_m=Q',\label{continm}
\end{equation}
\begin{equation}
\dot{\rho_r}+4H\rho_r=Q-Q',\label{continr}
\end{equation}
where $Q$ and $Q'$ stand for the interaction terms.

Taking a time derivative in both sides of Eq. (\ref{eqf1}), and
using Eqs. (\ref{Omega1}), (\ref{Omega2}), (\ref{eqf2}),
(\ref{continD}), (\ref{continm}), (\ref{continr}) and
$\xi=\varepsilon\rho_D H^{-1}$, the EoS parameter of interacting
viscous DE can be obtained as
\begin{equation}
\omega_D=-\frac{1}{3\Omega_D}\left\{2\left(\frac{\dot{H}}{H^2}-\Omega_k\right)\left[1-\left(\frac{\alpha}{2}\right)
\frac{\Omega_\alpha+(Hr_c)^{-2}}{1+\Omega_k}\right]
+3\Omega_m+4\Omega_r\right\}+3\varepsilon-1.\label{wlambda}
\end{equation}
The deceleration parameter is given by
\begin{equation}
q=-\left(1+\frac{\dot{H}}{H^2}\right).\label{q1}
\end{equation}
Substituting the term $\dot{H}/H^2$ from (\ref{wlambda}) into
(\ref{q1}) yields
\begin{equation}
q=\frac{(1+\Omega_k)}{2\left[1+\Omega_k-\frac{\alpha}{2}\Big(\Omega_\alpha+(Hr_c)^{-2}\Big)\right]}\Big[3\Omega_D(1+\omega_D-3\varepsilon)
+3\Omega_m+4\Omega_r\Big]-(1+\Omega_k).\label{q2}
\end{equation}
Using Eq. (\ref{eqf2}) one can rewrite (\ref{q2}) as
\begin{equation}
q=\frac{(1+\Omega_k)}{2\left[1+\Omega_k-\frac{\alpha}{2}\Big(\Omega_\alpha+(Hr_c)^{-2}\Big)\right]}\Big[1+\Omega_{k}+\Omega_{\alpha}(\alpha-3)
+\alpha(Hr_c)^{-2}+3\Omega_D(\omega_D-3\varepsilon)+\Omega_r\Big].\label{q3}
\end{equation}
%-----------------------------------------------------------------------------------------------
\section{The effect of the power-law correction term to the entropy
on the dynamics of DE}

Here to see how the power-law correction term to the entropy
(\ref{ec}) influence the dynamics of DE in our selected model for
the universe, we need to incorporate a specific form of the DE model
as well as the interaction terms between DE, DM and radiation. To do
this we consider the power-law entropy-corrected version of the
holographic DE (HDE) model. The HDE model is motivated by the
holographic principle \cite{Hooft}. Following \cite{Li}, the HDE
density is given by
\begin{equation}
\rho_{D}=3c^2M^2_PL^{-2},\label{HDE}
\end{equation}
where $c$ is a dimensionless constant, $M_P$ is the reduced Planck
Mass $M_P^{-2}=8\pi$ with $G=1$ and $L$ is the IR cut-off.

Indeed, the definition and derivation of the HDE density depends on
the Bekenstein-Hawking entropy-area relation $S_{\rm BH} = A/4$,
where $A\sim L^2$ is the area of horizon. Taking into account the
power-law correction (\ref{ec}) to the Bekenstein-Hawking entropy,
which appears in dealing with the entanglement of quantum fields
between in and out the horizon, the HDE density is modified
accordingly. This modification yields the energy density of the
so-called ``power-law entropy-corrected HDE'' (PLECHDE) as
\cite{sheyjam}
\begin{equation}\label{rhoPLECHDE}
\rho _{D }=3c^2M_{P}^{2}L^{-2}-\beta M_{P}^{2}L^{-\alpha},
\end{equation}
where $\beta$ is a dimensional constant. In the special case
$\beta=0$, the above equation yields the well-known HDE density
(\ref{HDE}).

From definition $\rho_{D}=3M_P^2H^2\Omega_{D}$ and using Eq.
(\ref{rhoPLECHDE}), we get
\begin{equation}
L=\frac{c}{H}\left(\frac{\gamma_c}{\Omega_{D}}\right)^{1/2},\label{HLPLECHDE}
\end{equation}
where
\begin{eqnarray}
\gamma_c = 1 - \frac{\beta}{3c^2}L^{2-\alpha}.\label{gammacPLECHDE}
\end{eqnarray}
If we consider the apparent horizon as an IR cut-off,
$L=\tilde{r}_A$, in a non-flat FRW universe, then taking a
derivative of Eq. (\ref{rhoPLECHDE}) with respect to cosmic time $t$
yields
\begin{equation}
\frac{\dot{\rho}_D}{\rho_D}=\left(\frac{\alpha-2}{\gamma_c}-\alpha\right)\frac{\dot{\tilde{r}}_A}{\tilde{r}_A}.
\label{rhodotPLECHDE1}
\end{equation}
Taking a time derivative of Eq. (\ref{ra}) gives
\begin{equation}
\dot{\tilde{r}}_A=\frac{\Omega_k-\frac{\dot{H}}{H^2}}{(1+\Omega_k)^{3/2}}.\label{radotPLECHDE}
\end{equation}
Taking a time derivative of Eq. (\ref{eqf1}) and using Eqs.
(\ref{Omega2}), (\ref{continm}) and (\ref{continr}) gives
\begin{equation}
\frac{\dot{H}}{H^2}=\Omega_k+\frac{\frac{8\pi}{3H^3}(\dot{\rho}_D+Q)-3\Omega_m-4\Omega_r}{2-\alpha\Big(\frac{\tilde{r}_{\rm
A}}{\tilde{r}_{\rm A_0}}\Big)^{2-\alpha}},\label{HdotH2PLECHDE}
\end{equation}
where following \cite{Saurya,Dvali} we take $r_c=\tilde{r}_{\rm
A_0}$. In what follows, following Cruz et al. in \cite{triple} we
assume
\begin{equation}
Q=3b^2H(\rho_D+\rho_m+\rho_r),\label{Q1}
\end{equation}
\begin{equation}
Q'=3b'^2H(\rho_D+\rho_m+\rho_r),\label{Q2}
\end{equation}
with the coupling constants $b^2$ and $b'^2$.

Using Eqs. (\ref{radotPLECHDE}), (\ref{HdotH2PLECHDE}) and
(\ref{Q1}) one can rewrite Eq. (\ref{rhodotPLECHDE1}) as
\begin{equation}
\frac{\dot{\rho}_D}{3H\rho_D}=\frac{\Big(\frac{\alpha-2}{\gamma_c}-\alpha\Big)\Big[\Omega_m+\frac{4}{3}\Omega_r-b^2(1+\Omega_k-\Omega_{\alpha})\Big]}
{\Big(\frac{\alpha-2}{\gamma_c}-\alpha\Big)\Omega_D+\Big[2-\alpha\Big(\frac{\tilde{r}_A}{\tilde{r}_{A_0}}\Big)^{2-\alpha}
\Big](1+\Omega_k)}.\label{rhodotPLECHDE2}
\end{equation}
Substituting Eqs. (\ref{Q1}) and (\ref{rhodotPLECHDE2}) in
(\ref{HdotH2PLECHDE}) gives
\begin{eqnarray}
\frac{\dot{H}}{H^2}=\Omega_k+\frac{3b^2(1+\Omega_k-\Omega_{\alpha})-3\Omega_m-4\Omega_r
}{\Big(\frac{\alpha-2}{\gamma_c}-\alpha\Big)\Big(\frac{\Omega_D}{1+\Omega_k}\Big)+\Big[2-\alpha\Big(\frac{\tilde{r}_{\rm
A}}{\tilde{r}_{\rm A_0}}\Big)^{2-\alpha}\Big]}
.\label{HdotH2PLECHDE2}
\end{eqnarray}
Inserting Eq. (\ref{HdotH2PLECHDE2}) in (\ref{wlambda}) and using
Eq. (\ref{Omega2}) yields the EoS parameter of the interacting
viscous PLECHDE as
\begin{eqnarray}
\omega_D=-1+3\varepsilon-b^2\left(\frac{1+\Omega_k-\Omega_{\alpha}}{\Omega_D}\right)
~~~~~~~~~~~~~~~~~~~~~~~~\nonumber\\
+\frac{\Big(\frac{\alpha-2}{\gamma_c}-\alpha\Big)\Big[b^2(1+\Omega_k-\Omega_{\alpha})-\Omega_m-\frac{4}{3}\Omega_r\Big]}
{\Big(\frac{\alpha-2}{\gamma_c}-\alpha\Big)\Omega_D+\Big[2-\alpha\Big(\frac{\tilde{r}_{\rm
A}}{\tilde{r}_{\rm
A_0}}\Big)^{2-\alpha}\Big](1+\Omega_k)}.\label{wDPLECHDE}
\end{eqnarray}
Replacing Eq. (\ref{HdotH2PLECHDE2}) into (\ref{q1}) gives the
deceleration parameter as
\begin{eqnarray}
q=-1-\Omega_k-\frac{3b^2(1+\Omega_k-\Omega_{\alpha})-3\Omega_m-4\Omega_r
}{\Big(\frac{\alpha-2}{\gamma_c}-\alpha\Big)\Big(\frac{\Omega_D}{1+\Omega_k}\Big)+\Big[2-\alpha\Big(\frac{\tilde{r}_{\rm
A}}{\tilde{r}_{\rm A_0}}\Big)^{2-\alpha}\Big]}.\label{qPLECHDE}
\end{eqnarray}
In the absence of correction term ($\alpha=0=\beta$), from Eqs.
(\ref{Omega2}) and (\ref{gammacPLECHDE}) we have $\Omega_{\alpha}=0$
and $\gamma_c=1$, respectively. If we also consider a spatially flat
FRW universe ($\Omega_k=0$), Eq. (\ref{ra}) shows that the apparent
horizon is same as the Hubble horizon, i.e. $\tilde{r}_A=H^{-1}$,
and $L=\tilde{r}_A=H^{-1}$. Now if we take
$\varepsilon=b^2=\Omega_r=0$ then Eq. (\ref{wDPLECHDE}) yields the
pressureless DE, i.e. $\omega_{D}=0$, where its EoS behaves like the
dust (or dark) matter. This result has been already obtained by Hsu
\cite{Hsu} for the HDE model with the IR cut-off $L=H^{-1}$. Also
from Eq. (\ref{qPLECHDE}) we obtain $q=1/2$. Therefore, choosing the
Hubble horizon as the IR cut-off $L=H^{-1}$ for the HDE model yields
a wrong EoS parameter and cannot drive the universe to accelerated
expansion.

Whereas in the presence of the power-law correction term, for the
flat FRW universe with $L=\tilde{r}_A=H^{-1}$ from Eq.
(\ref{HLPLECHDE}) we have $\gamma_c=\Omega_D/c^2$. Now if we take
$\varepsilon=b^2=\Omega_r=0$ then Eqs. (\ref{wDPLECHDE}) and
(\ref{qPLECHDE}) for the present time, $\tilde{r}_{\rm
A}=\tilde{r}_{\rm A_0}$, reduce to
\begin{equation}
\omega_{D_0}=-1-\left(\frac{1}{\Omega_{D_0}}-1\right)\left(1-\frac{1}{1-c^2-\Big(\frac{\alpha}{2-\alpha}\Big)\Omega_{D_0}}\right),
\label{omegaD0}
\end{equation}
\begin{equation}
q_0=-1+\frac{3(1-\Omega_{D_0})}{(2-\alpha)(1-c^2)-\alpha\Omega_{D_0}},\label{q0}
\end{equation}
where the subscript ``0'' denotes the present values of the
quantities. Taking $\Omega_{D_0}=0.73$ \cite{Riess} and $c=0.818$
\cite {Li6} then Eqs. (\ref{omegaD0}) and (\ref{q0}) yield
\begin{equation}
\omega_{D_0}=-1.021+\frac{0.480}{0.624-\alpha},\label{omegaD00}
\end{equation}
\begin{equation}
q_0=-\left(\frac{\alpha+0.139}{\alpha-0.624}\right).\label{q00}
\end{equation}
The above relations show that for $\alpha>0.624$ we have
$\omega_{D_0}<-1$ and $q_0<0$. Here $\omega_{D_0}<-1$ clears that
the PLECHDE with the IR cut-off $L=H^{-1}$ behaves like phantom DE.
Recent astronomical data indicates that the EoS parameter
$\omega_{D_0}$ at the present lies in a narrow strip around
$\omega_{D_0} = -1$ and is quite consistent with being below this
value \cite{Komatsu}. On the other hand, the satisfaction of the GSL
of gravitational thermodynamics for the universe with the power-law
corrected entropy (\ref{ec}) implies that $\alpha<2$ (see Appendix
A). Hence for the PLECHDE model with $0.624<\alpha<2$, the
identification of IR cut-off with the Hubble horizon $L=H^{-1}$, can
derive a phantom accelerating universe which is compatible with the
observations.
%-----------------------------------------------------------------------------------------------
\section{Conclusions}
Here, we considered a power-law quantum correction to the entropy of
the dynamical apparent horizon motivated by the entanglement of
quantum fields between inside and outside of the horizon. Using the
Clausius relation we obtained the modified Friedmann equation. For a
non-flat modified FRW universe filled with an interacting viscous DE
with DM and radiation, we obtained the EoS parameter of interacting
viscous DE as well as the deceleration parameter. We studied the
effect of the power-law correction term to the entropy on the
dynamics of DE in our selected model for the universe. Interestingly
enough, we found that for the PLECHDE model which is the power-law
entropy-corrected version of the HDE model, the identification of IR
cut-off with Hubble horizon, $L=H^{-1}$, can lead to a phantom
accelerating universe. This is in contrast to the ordinary HDE where
$\omega_D=0$ if one chooses $L=H^{-1}$. Furthermore, we investigated
the validity of the GSL of thermodynamics on the apparent horizon.
We found out that the GSL of thermodynamics with power-law corrected
entropy-area relation $S_{\rm A}=\frac{A}{4}\left[1-K_{\alpha}
A^{1-\frac{\alpha}{2}}\right]$ is satisfied for $\alpha<2$.
%-----------------------------------------------------------------------------------------------
\\
\\
\noindent{\textbf{Acknowledgements}\\ The authors thank the
anonymous referee for a number of valuable suggestions. The authors
also thank Professor Rabin Banerjee for useful discussions. The work
of K. Karami has been supported financially by Research Institute
for Astronomy and Astrophysics of Maragha, Iran.
%-----------------------------------------------------------------------------------------------
\begin{appendix}
\section{GSL with power-law corrected entropy}
Here, we study the validity of the GSL of thermodynamics for the
power-law entropy-corrected Friedmann equation. According to the
GSL, entropy of the viscous DE, DM and radiation inside the horizon
plus the entropy associated with the horizon must not decrease in
time.

Taking a time derivative in both sides of Eq. (\ref{ra}), and using
Eqs. (\ref{econs}), (\ref{eqf1}), (\ref{Omega1}), (\ref{Omega2}),
(\ref{eqf2}), (\ref{rho}) and (\ref{p2}), one can get
\begin{equation}
{\dot{\tilde{r}}_{\rm
A}}=\frac{(1+\Omega_k)^{-1/2}}{2\left[1+\Omega_k-\frac{\alpha}{2}\Big(\Omega_\alpha+(Hr_c)^{-2}\Big)\right]}\Big[3\Omega_D(1+\omega_D-3\varepsilon)
+3\Omega_m+4\Omega_r\Big].\label{radot1}
\end{equation}
Using Eq. (\ref{q2}) one can rewrite (\ref{radot1}) as
\begin{equation}
{\dot{\tilde{r}}_{\rm
A}}=\frac{1+\Omega_k+q}{(1+\Omega_k)^{3/2}}.\label{radot2}
\end{equation}
From Eqs. (\ref{ec}) and (\ref{TA1}), the evolution of the apparent
horizon entropy is obtained as
\begin{equation}
T_{\rm A}\dot{S}_{\rm A}=4\pi H{\tilde{r}}_{\rm
A}^3(\rho+p)-2\pi{\tilde{r}}_{\rm A}^2{\dot{\tilde{r}}}_{\rm
A}(\rho+p).\label{em}
\end{equation}
The entropy of the universe including the viscous DE, DM and
radiation inside the dynamical apparent horizon can be related to
its energy and pressure in the horizon by Gibb's equation
\cite{Izquierdo2}
\begin{equation}
T{{\rm d}S}={\rm d}(\rho V)+p~{\rm d}V=V{\rm d}\rho+(\rho+p){\rm
d}V,\label{Gibbs}
\end{equation}
where ${\rm V}=4\pi \tilde{r}_{\rm A}^3/3$ is the volume of the
universe enclosed by the dynamical apparent horizon $\tilde{r}_{\rm
A}$. Following \cite{Sheykhi2,Karami1}, we assume that the
temperature $T$ of the universe enclosed by the dynamical apparent
horizon should be in equilibrium with the Hawking temperature
$T_{\rm A}$ associated with the dynamical apparent horizon, so we
have $T = T_{\rm A}$. Therefore from Eq. (\ref{Gibbs}) one can
obtain
\begin{equation}
T_{\rm A}\dot{S}=4\pi{\tilde{r}}_{\rm A}^2{\dot{\tilde{r}}_{\rm
A}}(\rho+p) -4\pi H{\tilde{r}}_{\rm A}^3(\rho+p),\label{en}
\end{equation}
where $S=S_D+S_m+S_r$ is the entropy in the universe containing the
viscous DE, DM and radiation. Finally, adding  Eqs. (\ref{em}) and
(\ref{en}), the GSL due to the different contributions of the
viscous DE, DM, radiation and dynamical apparent horizon can be
obtained as
\begin{equation}
T_{\rm A}\dot{S}_{\rm {tot}}=2\pi{\tilde{r}}_{\rm
A}^2{\dot{\tilde{r}}_{\rm A}}(\rho+p),\label{eu}
\end{equation}
where $S_{\rm tot}=S+S_{\rm A}$ is the total entropy.

From Eqs. (\ref{rho}), (\ref{p2}) and using (\ref{Omega1}) one can
obtain
\begin{equation}
\rho+p=\frac{H^2}{8\pi}\Big[3\Omega_D(1+\omega_D-3\varepsilon)
+3\Omega_m+4\Omega_r\Big].\label{rhop1}
\end{equation}
Using Eq. (\ref{q2}) one can rewrite (\ref{rhop1}) as
\begin{equation}
\rho+p=\frac{H^2}{4\pi
}\left(\frac{1+\Omega_k+q}{1+\Omega_k}\right)\left[1+\Omega_k-\frac{\alpha}{2}\left(\Omega_\alpha+(Hr_c)^{-2}\right)\right].\label{rhop2}
\end{equation}
Substituting Eqs. (\ref{ra}), (\ref{radot2}) and (\ref{rhop2}) into
(\ref{eu}) yields the GSL as
\begin{equation}
T_{\rm A}\dot{S}_{\rm
{tot}}=\frac{(1+\Omega_k+q)^2}{2(1+\Omega_{k})^{7/2}}\left[1+\Omega_k-\frac{\alpha}{2}\left(\Omega_\alpha+(Hr_c)^{-2}\right)\right],\label{GSL3}
\end{equation}
which can be rewritten by the help of Eq. (\ref{q2}) as
\begin{equation}
T_{\rm A}\dot{S}_{\rm
{tot}}=\frac{\Big[3\Omega_D(1+\omega_D-3\varepsilon)
+3\Omega_m+4\Omega_r\Big]^2}{8(1+\Omega_{k})^{3/2}\left[1+\Omega_k-\frac{\alpha}{2}\Big(\Omega_\alpha+(Hr_c)^{-2}\Big)\right]}.\label{GSL4}
\end{equation}
By substituting $\Omega_\alpha$ from Eq. (\ref{Omega2}) in the above
equation we get
\begin{equation}
T_{\rm A}\dot{S}_{\rm
{tot}}=\frac{\Big[3\Omega_D(1+\omega_D-3\varepsilon)
+3\Omega_m+4\Omega_r\Big]^2}{8(1+\Omega_{k})^{5/2}\left[1-\frac{\alpha}{2}\left(\frac{\tilde{r}_{\rm
A}}{r_c}\right)^{2-\alpha}\right]}.\label{GSL5}
\end{equation}
According to Eq. (\ref{GSL5}), the validity of GSL, i.e. $T_{\rm
A}\dot{S}_{\rm {tot}}>0$, depends on the sign of the expression
$\left[1-\frac{\alpha}{2}\left(\frac{\tilde{r}_{\rm
A}}{r_c}\right)^{2-\alpha}\right]$ appearing in the denominator.
Hence the GSL is hold when
\begin{equation}
1-\frac{\alpha}{2}\left(\frac{\tilde{r}_{\rm
A}}{r_c}\right)^{2-\alpha}=1-\frac{\alpha}{2}\left(\frac{\tilde{r}_{\rm
A}}{\tilde{r}_{{\rm A}_0}}\right)^{2-\alpha}>0,\label{con}
\end{equation}
where following \cite{Saurya,Dvali} we identify the crossover scale
$r_c$ with the present value of the apparent horizon
$\tilde{r}_{{\rm A}_0}$. Since $\tilde{r}_{\rm A}/\tilde{r}_{{\rm
A}_0}$ tends to zero in the far past and its today value is
$\tilde{r}_{\rm A}/\tilde{r}_{{\rm A}_0}=1$, hence for $0\leq
\tilde{r}_{\rm A}/\tilde{r}_{{\rm A}_0}\leq 1$ the condition
(\ref{con}) is satisfied when $\alpha<2$. Therefore the GSL with the
power-law corrected entropy (\ref{ec}) is respected for $\alpha<2$.
\end{appendix}
%-----------------------------------------------------------------------------------------------

%------------------------------------------------------------------------------------------------
\end{document}